\documentclass[prd,aps,reprint,onecolumn,superscriptaddress,tightenlines,nofootinbib,eqsecnum,preprintnumbers,longbibliography,12pt]{revtex4-2}

\usepackage{epsfig,amsfonts,mathrsfs,amsmath,amssymb,graphicx,color,slashed,multirow}
\usepackage{amsmath,latexsym,amssymb,graphicx,slashed,hyperref,color,enumerate,url,etoolbox,cancel,gensymb,physics}
\hypersetup{colorlinks,citecolor= nicegreen,linkcolor= nicered}
\usepackage[usenames,dvipsnames]{xcolor}
\definecolor{nicered}{rgb}{0.7,0.1,0.1}
\definecolor{nicegreen}{rgb}{0.1,0.5,0.1}
\definecolor{niceblue}{rgb}{0.1,0.2,0.6}

\usepackage{bbm}
\usepackage{subfig}
\usepackage{physics}
\usepackage{dsfont}
\begin{document}
\def\Carleton{Ottawa-Carleton Institute for Physics, Carleton University, Ottawa, ON K1S 5B6, Canada}
\title{Status of negative coupling modifiers for extended Higgs sectors}
\author{Carlos H. de Lima}
\author{Daniel Stolarski}
\affiliation{\Carleton}
\author{Yongcheng Wu}
\affiliation{Department of Physics, Oklahoma State University, Stillwater, OK, 74078, USA}
\begin{abstract}
In this work, we study the status of negative coupling modifiers in extended Higgs sectors, focusing on the ratio of coupling modifiers that probes custodial symmetry violation $\lambda_{WZ} = \kappa_{W}/\kappa_{Z}$. Higgs sectors with multiplets larger than doublets are the only weakly coupled models that give tree-level modifications to $\lambda_{WZ}$, and we explore all such models allowed by the constraint from the $\rho$ parameter and perturbative unitarity. This class of models has a custodial symmetry violating potential, while the vacuum configuration preserves the symmetry. We apply precision measurements from ATLAS and CMS and show that each data set can exclude a vast set of models with $\lambda_{WZ} < 0$ at greater than 95\% confidence level. We give evidence that $\lambda_{WZ}<0$ is excluded in all weakly coupled models.

 [\ref{errata} for this work on page 19 changing our claims of complete exclusion of negative $\lambda_{WZ}$ to a softer version where only the minimal models are excluded.]
\end{abstract}
\maketitle

\section{Introduction}

The Higgs boson was the last missing piece for the standard model (SM). After its discovery \cite{higgsDISC1,higgsDISC2}, the hunt to uncover its underlying proprieties started. The LHC experiments have measured many of its properties and showed that the corrections from new physics, if they exist, appear to be small~\cite{ATLAS:2016neq,CMSlwz,ATLASlwz}. We know, however, the SM cannot be a complete theory of nature, so a thorough search for deviations in the Higgs sector is of paramount importance.

The fact that deviations from the properties of the SM Higgs appear to be small may be an artifact of the way we are accessing its information. Most observables of the Higgs sector are cross-sections and decay rates, which normally are not sensitive to the sign of the underlying coupling. This means that some new physics could hide in plain sight if it generates couplings accidentally close to the SM, but with flipped signs. A simple example of such a scenario can be explored in looking at modifications of the Higgs couplings to electroweak gauge bosons. If we parameterize deviations from the SM predictions to those couplings using $\kappa_{W}$ and $\kappa_{Z}$~\cite{LHCHiggsCrossSectionWorkingGroup:2013rie}, then the agreement of the measurements of $H\rightarrow ZZ^*$~\cite{ATLAS:2017qey,CMS:2021ugl} and $H\rightarrow WW^*$~\cite{ATLAS:2018xbv,CMS:2020dvg} with the SM prediction indicate that $\kappa_{W}^2 \approx 1$ and $\kappa_{Z}^2 \approx 1$. If we define
\begin{align}
\lambda_{WZ} =\frac{\kappa_{W}}{\kappa_{Z}} \, ,
\label{eq:lamwz}
\end{align}
then the data imply $\lambda_{WZ} \approx \pm 1$, but these processes cannot distinguish the negative from the positive sign scenario. This can be confirmed with a global fit of all the Higgs data~\cite{CMSlwz,ATLASlwz}, which indicates that $|\lambda_{WZ}| \approx 1$ with $\mathcal{O}(10\%)$ precision. The ATLAS analysis~\cite{ATLASlwz}, however, assumes $\lambda_{WZ} >0$ in its fit. The CMS analysis~\cite{CMSlwz} does allow either sign, but it has almost no discrimination power for the sign. Interestingly, a negative value of $\lambda_{WZ}$ is slightly preferred in the CMS fit.

Higgs couplings with similar magnitudes but opposite signs of the SM prediction can be probed using interference effects, for example in Higgs decays to four leptons~\cite{Chen:2016ofc}, $W^+W^- H$ production~\cite{Chiang:2018fqf}, VBF-VH production~\cite{Stolarski:2020qim}, and the combination of $Zh$ and $tH$ production\cite{Xie:2021xtl}. These measurements, however, could also be affected by the presence of new states that can also contribute to the interference measured, so it is difficult to make a model-independent determination of the sign of the couplings.

A particularly interesting feature of the Higgs sector in the SM is that it exhibits an accidental custodial symmetry~\cite{CUSTODIAL1} of $SU(2)_L\times SU(2)_R$. This custodial symmetry predicts the value of the $\rho$ parameter:
\begin{align}
\rho = \frac{m_{W}^{2}}{c_{w}^{2}m_{Z}^{2}} \approx 1\, ,
\end{align}
After a more careful accounting of custodial symmetry breaking effects within the SM, the measured value $\rho$ parameter agrees with the SM prediction with a precision $\sim 10^{-4}$~\cite{ParticleDataGroup:2020ssz}. This is very strong evidence that custodial symmetry is realized in nature. The custodial symmetry also predicts $\lambda_{WZ} = 1$, and taking the precision of the $\rho$ parameter as a naive guide, deviations from 1 of $\lambda_{WZ}$ are expected to be small.

In order to quantitatively explore these flipped sign scenarios, we must have models that go beyond the SM and can modify the Higgs couplings. We are particularly interested in the $\lambda_{WZ}<0$ scenario, which means that these models must break custodial symmetry.\footnote{A custodial 5-plet does have $\lambda_{WZ}=-1/2$~\cite{Low:2010jp}, but such a state has no couplings to fermions, so it is highly implausible that the $H(125)$ is a 5-plet.} The only weakly coupled models that give tree-level modifications of $\lambda_{WZ}$ are models with extended Higgs sectors that have representations larger than doublets, which are the models we focus on in this work. These models are of course strongly constrained by the measurement of the $\rho$ parameter. This constraint can be evaded in models that have custodial symmetry built into them such as the Georgi-Machacek model~\cite{GM1,GM2} and its generalizations~\cite{GGM}.  In those models, the vevs of scalars in the same custodial representation are equal and their contributions to the $\rho$ parameter cancel at tree level. In that case, however, modifications to $\lambda_{WZ}$ are also small.

In this work, we highlight that the $\rho$ parameter has its source in the vacuum configuration of the extended scalar sector, while the Higgs coupling modifiers have most of their contributions coming from the mixing between multiplets. This means that there are models that avoid the $\rho$ parameter constraint and can still have a large amount of custodial violation. These models have a vacuum expectation value (vev) that is custodial symmetric, but a potential that violates such symmetry. We call this class of models accidentally custodial symmetric (AC). These models have the same field content as generalized Georgi-Machacek models~\cite{GGM}, while having a custodial breaking potential. For completeness, we will also consider more general field contents where contributions to the $\rho$ parameter from different custodial multiplets cancel one another.\footnote{The contributions from different multiplets are independent of one another, so having their effects cancel is of course a fine-tuning, but in this work we seek to explore even fine-tuned models.
}

Naively, one would think that it is necessary to work out the potential for each possible extension and then perform individual parameter scans, which would make it computationally infeasible to systematically cover the whole parameter space. We can make the following observation to avoid this problem: the parameters $\rho$ and $\lambda_{WZ}$ depend only on the Higgs vevs and the matrix that rotates between the gauge and mass bases for the Higgs states. While the vevs and rotation matrices do in turn depend on the full scalar potential of the model, we will assume that they are independent and consider the most general possibility. Since we can then treat both the Higgs eigenvector and the vev as independent quantities, the parameter scan is lower-dimensional compared to scanning the potential, and we can explore a wider class of models. We will then show that even with this assumption, all models studied are excluded at more than $95\%~\text{CL}$ using the ATLAS or CMS data.\footnote{Using the ATLAS analysis, one model is excluded at $99.5\%$ CL, while all the rest are excluded at more than 99.7\% CL (3$\sigma$).
The CMS analysis does not give a correlation matrix for the ratio of coupling modifiers making it difficult to make a more precise statement than we make here.} Given this exclusion, it becomes unnecessary to study the full potential of the model, since any solution arising from the potential will be covered by our analysis and thus excluded.

The remainder of this work is organized as follows. In section~\ref{sec:ACT}, we introduce the AC triplet scenario which is the simplest one that can give rise to large modifications of $\lambda_{WZ}$ and be consistent with bounds from the $\rho$ parameter.  In section~\ref{sec:ACGEN}, we generalize to all multiplets allowed by perturbative unitarity and discuss the set of models that we are considering. In section~\ref{sec:EXP}, we apply the experimental bounds to the models and highlight the general features that are common among all of them. We conclude in section~\ref{sec:CONC}, and various technical details are given in the appendices.

\section{Accidentally custodial symmetric triplets} \label{sec:ACT}

In extended electroweak sectors, particularly those with scalar representations larger than doublets, the custodial violation can come from two distinct sources: the vacuum configuration, and the Lagrangian. The vacuum contribution modifies the $\rho$ parameter at tree-level and it is heavily constrained \cite{ParticleDataGroup:2020ssz}. The custodial violation from  interactions enters only at loop-level for the $\rho$ parameter and is thus less constrained.  In the Standard Model, this is the case of the hypercharge and Yukawa breaking of custodial symmetry. We are interested in models which can avoid the $\rho$ parameter bound and still have large custodial violation. The class of models that satisfy these conditions have a custodial violating potential, but they have a limit where the vacuum is custodial symmetric.

The simplest such model, which we study in this section, has the same field content as the Georgi-Machacek model~\cite{GM1,GM2}, while the potential is the most general allowed by the standard model symmetries. We have the usual Higgs doublet ($\phi^{+}$, $\phi^{0}$) with hypercharge $Y=1$\footnote{The quark doublet has $Y=1/3$ in this convention.}, a complex triplet ($\chi^{++}$, $\chi^{+}$, $\chi^{0}$) with hypercharge $Y=2$ and a real triplet ($\xi^{+}$, $\xi^{0}$, $\xi^{-}$) with hypercharge $Y=0$. The most general potential can be written as:
\begin{align}
V &= \mu_{2}^{2} \phi^{\dagger}\phi  +
\mu_{3}^{\prime\,2} \chi^{\dagger}\chi +
 \frac{\mu_{3}^{2}}{2} \xi^{\dagger}\xi  + \lambda_{1} \left( \phi^{\dagger}\phi \right)^{2} +\lambda_{2} |\chi^{C \dagger} \chi |^{2} + \, \nonumber\\
&+ \lambda_{3} (\phi^{\dagger} \tau^{a} \phi) (\chi^{\dagger} t^{a}\chi) + \lambda_{4}\left[ (\phi^{C \dagger}\tau^{a}\phi)(\chi^{\dagger} t^{a}\xi) + \text{h.c}  \right] +  \nonumber\\
&+ \lambda_{5} (\phi^{\dagger}\phi)(\chi^{\dagger}\chi) + \lambda_{6}(\phi^{\dagger}\phi) (\xi^{\dagger}\xi) + \lambda_{7} (\chi^{\dagger}\chi)^{2} + \lambda_{8}(\xi^{\dagger}\xi)^{2} + \nonumber\\
&+ \lambda_{9} |\chi^{\dagger} \xi|^{2} + \lambda_{10} (\chi^{\dagger}\chi) (\xi^{\dagger}\xi) - \frac{1}{2}\left[ M'_{1} \phi^{\dagger}\Delta_{2} \phi^{C} + \text{h.c} \right] + \nonumber\\
&+ \frac{M_{1}}{2}\phi^{\dagger} \Delta_{0}\phi - 6 M_{2} \chi^{\dagger} \bar{\Delta}_{0} \chi \, .
\label{eq:full-potential}
\end{align}
where $\tau_{i}$ and $t_{i}$ are the generators of the doublet and triplet representations that can be seen in Appendix \ref{ac:group}. It is also defined:
\begin{align}
\Delta_{2} &\equiv \begin{pmatrix}
\frac{\chi^{+}}{\sqrt{2}} & -\chi^{++} \\
\chi^{0} & -\frac{\chi^{+}}{\sqrt{2}}
\end{pmatrix} \, , \\
\Delta_{0} &\equiv \begin{pmatrix}
\frac{\xi^{0}}{\sqrt{2}} & -\xi^{+} \\
-\chi^{+*} & -\frac{\chi^{0}}{\sqrt{2}}
\end{pmatrix} \, , \\
\bar{\Delta}_{0} &\equiv \begin{pmatrix}
-\xi^{0} & \xi^{+} & 0 \\
\xi^{+*} & 0 & \xi^{+} \\
0 & \xi^{+*} & \xi^{0}
\end{pmatrix} \, ,
\end{align}
and the charge conjugation defined as:
\begin{align}
\phi^{C} \equiv \begin{pmatrix}
0 & 1 \\
-1 & 0
\end{pmatrix} \phi^{*}  \, , \\
\chi^{C} \equiv \begin{pmatrix}
0 & 0 & 1 \\
0 & -1 & 0 \\
1 & 0 & 0
\end{pmatrix} \chi^{*}
\end{align}

This is the most general renormalizable potential for these fields and was defined in \cite{CUSGM2,CUSGM1,CUSGM3} to study the custodial violation in the GM model from the loop corrections. In the custodial limit, $\chi$ and $\xi$ can be organized into a bi-triplet written as:
\begin{align}
X=  \begin{pmatrix}
\chi^{0 *} & \xi^{+} & \chi^{++} \\
-\chi^{+*} & \xi^{0} & \chi^{+} \\
\chi^{++*} & -\xi^{+*} & \chi^{0} \, ,
\end{pmatrix}
\end{align}
and the allowed couplings can all be written in terms of $X$.
The specific relations between the couplings that enforce the custodial limit can be seen in Eq.~(27) of \cite{CUSGM1}. In our case, we assume that the custodial symmetry is not an underlying symmetry of the potential, rather it is emergent from the vacuum configuration. Since the $\rho$ parameter at tree-level is mostly sensitive to the vacuum, we can have a large custodial violation without large contributions to the $\rho$ parameter.
One loop corrections do contribute to the $\rho$ parameter, and even if they are small, they can be of the same order as the experimental precision. This will generate additional bounds on the parameters of the model~\cite{CUSGM2}. Since we want to study the custodial vacuum one should also determine whether such a configuration is stable. This analysis can be done using the methods developed in~\cite{VACSTABILITY,Z2VAC}. In the analysis that we will perform here, these bounds can enter in the final stages, if there is any parameter space left.

The couplings of the scalars to the gauge bosons come from the kinetic term that has the standard form:
\begin{align}
\mathcal{L}_{\text{kin}} =  \left( D_{\mu}\phi \right)^{\dagger} D_{\mu} \phi  + \left( D_{\mu}\chi \right)^{\dagger}D_{\mu}\chi + \frac{1}{2} \left( D_{\mu}\xi \right)^{\dagger}D_{\mu}\xi \, ,
\end{align}
where the covariant derivative depends on the representation of the field:
\begin{align}
D_{\mu} = \partial_{\mu} - \frac{i g}{\sqrt{2}} (W^{-}t_{+}+W^{+}t_{-}) - \frac{i e}{s_{w}c_{w}} (t_{3}-s_{w}^{2}Q) \, .
\end{align}
The basis for each representation for $t_{i}$ can be found in the Appendix \ref{ac:group}, the charge matrix and $t^{\pm}$ is defined as:
\begin{align}
Q &= t_{3} + Y/2 \, , \\
t_{\pm} &=  t_{1} \mp i t_{2} \, .
\end{align}
Couplings of the scalars to fermions are very similar to the SM: in the gauge basis, the doublet couples to all fermions while the triplets do not.

We are interested here in the contributions from this model to the coupling modifiers of the Higgs. After electroweak (EW) symmetry breaking, we have the following field redefinitions for the  vacua:
\begin{align}
\phi^{0} &= \frac{\nu_{\phi}}{\sqrt{2}} + \frac{1}{\sqrt{2}} ( \phi^{0}_{R}+i\phi^{0}_{I}) \, ,  \;\;\;  \xi^{0} = \nu_{\xi} +\xi^{0}_{R} \, ,   \;\;\;
\chi^{0} = \nu_{\chi} + \frac{1}{\sqrt{2}} ( \chi^{0}_{R} + i \chi^{0}_{I} ) \, .
\end{align}
Each CP even neutral component has a coupling to gauge bosons. In this specific case, we have from the kinetic term:
\begin{align}
g_{\phi^{0}_{R}WW} &= \frac{g^{2}}{2} \nu_{\phi}  \, , \;\;\; g_{\phi^{0}_{R}ZZ} = \frac{e^{2}}{2c_{w}^{2}s_{w}^{2}}\nu_{\phi} \, , \nonumber\\
g_{\chi^{0}_{R}WW} &= \sqrt{2} g^{2} \nu_{\chi} \, , \;\;\ g_{\chi^{0}_{R}ZZ} = \frac{2\sqrt{2}e^{2}}{c_{w}^{2}s_{w}^{2}} \nu_{\chi} \, , \nonumber\\
\label{eq:gauge-couplings}
g_{\xi^{0}_{R}WW} &=  2 g^{2} \nu_{\xi} \, , \;\;\ g_{\xi^{0}_{R}ZZ} = 0 \, .
\end{align}
These couplings are in the gauge basis, but generically all the states with the same electric charge will mix. Therefore, once we go to the mass basis, states will couple with linear combinations of the couplings in Eq.~\eqref{eq:gauge-couplings}. One mass eigenstate will be the 125 GeV Higgs, and its overlap with the different gauge eigenstates can be used to compute the coupling to the gauge bosons. Measurements of the couplings of the Higgs at the LHC can then be used to constrain the model.

As noted in the introduction, deviations from the SM predictions of couplings can be parametrized in terms of coupling modifiers by dividing out the SM value~\cite{LHCHiggsCrossSectionWorkingGroup:2013rie}. To set the notation, we first define the total Standard Model vacuum expectation value in terms of the Fermi constant $G_{F}$:
\begin{align}
\nu = \left(\sqrt{2} G_{F} \right)^{-\frac{1}{2}} \approx 246 \,  \text{GeV} \, .
\label{eq:vGf}
\end{align}
Then, let us define the general coupling modifier for a scalar field $X$ as:
\begin{align}
\kappa_{W}^{X} = \frac{g_{XWW}}{\frac{\left(g^{2}\nu\right)}{2}} \, , \;\;\; \kappa_{Z}^{X} = \frac{g_{XZZ}}{\frac{\left(e^{2}\nu\right)}{2 c_{w}^{2}s_{w}^{2}}} \, .
\end{align}
We can include the fermion coupling modifiers by noting that the fermions only couple the doublet in the gauge basis. So in the gauge basis, the coupling modifiers are given by\footnote{The \ref{errata} addresses the wrong fermionic coupling modifiers in this equation and the modifications of the results derived in this work.}:
\begin{align} \label{eq:tag}
\kappa_{f}^{\phi} &=  \frac{\nu_{\phi} }{\nu} \, , \;\;\;  \kappa_{W}^{\phi} = \frac{\nu_{\phi} }{\nu} \, , \;\;\; \kappa_{Z}^{\phi} = \frac{\nu_{\phi} }{\nu}  \, , \nonumber\\
\kappa_{f}^{\chi} &= 0 \, , \;\;\;  \kappa_{W}^{\chi} = \frac{2\sqrt{2} \nu_{\chi}}{\nu}  \, , \;\;\; \kappa_{Z}^{\chi} = \frac{4\sqrt{2}\nu_{\chi}}{\nu} \,  \, , \nonumber\\
\kappa_{f}^{\xi} &= 0 \, , \;\;\;  \kappa_{W}^{\xi} = \frac{4\nu_{\xi}}{\nu}  \, ,\;\;\; \kappa_{Z}^{\xi} = 0 \, .
\end{align}
The coupling modifiers in the mass basis can then be computed in terms of the rotation matrix between the two bases.
This approach can be generalized to any multiplet. The expressions of different coupling modifiers from different multiplets can be seen in Appendix~\ref{ap:kappas}.

Finally we can work out the $\rho$ parameter at tree-level. The mass of the $W$ and $Z$ boson in this model is:
\begin{align}
m_{W}^{2} = \frac{g^{2}}{4} \left(\nu_{\phi}^{2} + 4 \nu_{\chi}^{2} + 4 \nu_{\xi}^{2} \right) \, , \\
m_{Z}^{2} = \frac{e^{2}}{4 c_{w}^{2}s_{w}^{2}} \left( \nu_{\phi}^{2} + 8 \nu_{\chi}^{2} \right) \, .
\end{align}
From the definition of the $\rho$ parameter we have:
\begin{align}
\rho = \frac{m_{W}^{2}}{c_{w}^{2} m_{Z}^{2}} = \frac{\nu_{\phi}^{2}+4\nu_{\chi}^{2}+4\nu_{\xi}^{2}}{\nu_{\phi}^{2}+8\nu_{\chi}^{2}} \, .
\label{eq:rho-triplet}
\end{align}
We can see that we restore $\rho = 1$ if $\nu_{\chi} = \nu_{\xi}$ as in the original GM model~\cite{GM1,GM2}. This is the custodial limit of this model. Given the precise measurement of $\rho$, the model is excluded unless $\nu_{\chi} \approx \nu_{\xi}$. For models with more than two multiplets, we can have configurations that cancel the contribution of the $\rho$ parameter without being a custodial symmetric vacuum, which we explore further in the analysis of the other models.

Now, we posit that the model is in the custodial vacuum and study the mixing between the gauge eigenstates. The important mixing is the one that will generate the 125 GeV Higgs eigenstate in the neutral sector. The mass matrix, in the basis $( \chi^{0}_{R},\xi^{0}_{R}, \phi^{0}_{R})$ in terms of the parameters of Eq.~\eqref{eq:full-potential} is:
\footnotesize{
\begin{align}
\mathcal{M}^{2}_{0} = \begin{pmatrix}
\nu_{\phi}^{2}\left( \frac{M'_{1}}{4\nu_{\chi}} - \frac{\lambda_{4}}{2\sqrt{2}} \right) + 4 \lambda_{7} \nu_{\chi}^{2}  & \nu_{\phi}^{2}\frac{\lambda_{4}}{2} + 2\sqrt{2}\lambda_{10}\nu_{\chi}^{2}-6\sqrt{2} M_{2} \nu_{\chi}  & \nu_{\phi}\nu_{\chi} \left( \lambda_{4} + \sqrt{2}\lambda_{5}+\frac{\lambda_{3}}{\sqrt{2}} \right) - M'_{1} \frac{\nu_{\phi}}{\sqrt{2}} \\
\nu_{\phi}^{2}\frac{\lambda_{4}}{2} + 2\sqrt{2}\lambda_{10}\nu_{\chi}^{2} - 6 \sqrt{2} M_{2} \nu_{\chi} & \nu_{\phi}^{2} \left( \frac{M_{1}}{4\nu_{\chi}} - \frac{\lambda_{4}}{\sqrt{2}}  \right) +8 \lambda_{8} \nu_{\chi}^{2}+ 6 M_{2} \nu_{\chi}  & \nu_{\phi}\nu_{\chi} \left( 2 \lambda_{6} +\sqrt{2}\lambda_{4} \right) - \frac{M_{1}}{2} \nu_{\phi} \\
\nu_{\phi}\nu_{\chi} \left( \lambda_{4} + \sqrt{2}\lambda_{5}+\frac{\lambda_{3}}{\sqrt{2}} \right) - M'_{1} \frac{\nu_{\phi}}{\sqrt{2}} & \nu_{\phi}\nu_{\chi} \left( 2 \lambda_{6} +\sqrt{2}\lambda_{4} \right) - \frac{M_{1}}{2} \nu_{\phi} & 2\lambda_{1} \nu_{\phi}^{2}
\end{pmatrix}
\end{align}}
\normalsize%

At this point, one could do a scan for the full parameter space of the model and obtain for each parameter point an eigenvector that corresponds to the Higgs. Here we take a different approach and consider the Higgs mass eigenvector and the set of vevs to be independent. This will encompass all possible model points, and may also contain points that are unphysical. Therefore, if we can exclude this more general parameterization, we can conclude that the model is indeed excluded.

The 125 GeV Higgs eigenvector in the most general form can be written as:
\begin{align}
h = R_{1} \chi_{R}^{0} + R_{2} \xi_{R}^{0} + R_{3} \phi_{R}^{0} \, .
\end{align}
The vevs can be written as:
\begin{align}
\nu^{2} = \nu_{\phi}^{2} + 4 \nu_{\chi}^{2} + 4\nu_{\xi}^{2} \, .
\label{eq:total-vev}
\end{align}
Let us now investigate the behaviour of the ratios of coupling modifiers for the AC custodial triplet ($\nu_{\chi} = \nu_{\xi}$). In terms of these parameters we have:
\begin{align}
\lambda_{WZ} = \frac{2 \sqrt{2} \nu_{\chi}R_{1}+4 \nu_{\chi}R_{2} +\nu_{\phi}R_{3}}{4 \sqrt{2} \nu_{\chi}R_{1}+\nu_{\phi}R_{3}} \, .
\end{align}
If we assume that $\lambda_{WZ} = -1$, we can find a relation between the vector components and the vevs:
\begin{align}
R_{1} = -\frac{R_{3}\nu_{\phi}+2\nu_{\chi} R_{2}}{3\sqrt{2}\nu_{\chi}} \,.
\end{align}
Now, we can explore if it is possible for the other coupling modifiers to be close to $\pm 1$ to be consistent with Higgs data. In this case we can use:
\begin{align}
\lambda_{fZ} =\frac{ \kappa_{f}}{\kappa_{Z}} \, ,\label{eq:lamfz}\\
\kappa_{fZ} = \frac{\kappa_{f}\kappa_{Z}}{\kappa_{h}} \label{eq:kapfz}\, ,
\end{align}
 where we use $\kappa_{h}$ parameterizes the deviation of the total width of the Higgs away from the SM value,
 \begin{equation}
 \Gamma_h = \frac{\kappa_h^2 \,\Gamma_h^\text{SM}}{1-B_\text{BSM}} \, .
 \label{eq:kap-h}
 \end{equation}
 Without the inclusion of loop induced processes,\footnote{The only limit where $\kappa_{\gamma}$ or  $\kappa_{Z\gamma}$ could give important contributions to $\kappa_{h}$ in a realistic scenario is where both $\kappa_{f}$ and $\kappa_{V}$ are small. However, in such a limit $\kappa_{fZ}$ behaves as $\kappa_{fZ} \sim 0$ which is excluded. We assume that the gluon coupling modifier is controlled by $\kappa_f$, since we do not include new colored states.\label{floop}} $\kappa_h$ can be written as:
\begin{align}
\kappa_{h} = \sqrt{0.75\kappa_{f}^{2} + 0.22\kappa_{W}^{2} +0.03 \kappa_{Z}^{^2}} \, .
\end{align}
First, let us assume that $\lambda_{fZ} = \pm 1$, this can give us the following relation:
\begin{align}
R_{2} = \mp \frac{R_{3}\nu_{\phi}}{2\nu_{\chi}} \, .
\end{align}
Finally, we want to know what are the possible values for the last observable $\kappa_{fZ}$, doing the substitutions we have:
\begin{align} \label{eq:kfZMOD}
\kappa_{fZ} = R_{3}\frac{\nu_{\phi}}{\nu} \, \, \text{for} \, \, \lambda_{fZ} =  1 \, , \\
\kappa_{fZ} \approx -1.39 R_{3}\frac{\nu_{\phi}}{\nu} \, \, \text{for} \, \, \lambda_{fZ} =  -1 \, .
\end{align}
From this result, we can see that it is impossible to have an accidental cancellation in $\lambda_{WZ}$ and still be close to the Standard Model value for the other coupling modifiers. We can at most have two of these three coupling modifiers close to the SM value. In Eq.~\eqref{eq:kfZMOD} both $R_{3}$ and $\frac{\nu_{\phi}}{\nu}$ are smaller than one, moving away from the SM value. Using the unitarity of $\vec{R}$ we can place an upper bound on of  $\kappa_{fZ} \lesssim 0.6$. This highlights that accidental cancellations can happen, but they can generally be constrained by multiple observables. This analysis does not include the correlation between $\nu_{\chi}$ and $\vec{R}$ which is present in the model and could even further constrain the parameter space.

Using this type of reasoning we can explore this region of parameter space without scanning the 16 parameters of the theory. In the end, if the model or a region of the parameter space is excluded for general vevs and eigenvectors, the correlations of the variables will not change the exclusion. When going to models with higher representations, this procedure will generalize to generating a random vector $\vec{R}$ and a random vev configuration.
  In summary, treating the vectors as independent can exclude some parts of the parameter space, but if some configurations are not excluded by the data, the model would need to be further investigated. This is the approach that we use for this model and the generalization that we introduce in the next section.

\section{General parametrization of AC extended scalar sectors}\label{sec:ACGEN}

We now generalize the analysis of the last section to larger representations of $SU(2)$. We already saw that the AC triplets have the possibility of negative couplings provided the mass eigenvector has the correct values. Any extended sector with multiplets larger than doublet can also have a region in parameter space where $\lambda_{WZ} \approx -1$, provided the potential is custodial violating.
 We will continue to use the nomenclature of accidentally symmetric (AC) for models that have a custodial limit satisfied by the vacuum but not the potential.
 The AC models can generate negative couplings in a simple way, but we could also have a situation where we are not in the symmetric vacuum or we do not have any custodial symmetric limit. We also investigate these possibilities to have an overall picture of the status of negative coupling modifiers for any extended Higgs sector. 

Since we want to give a general statement of the state of such negative couplings for an arbitrary extended sector, we first assess how many different models there are that can have $\lambda_{WZ} \approx -1$. We can use perturbative unitarity of boson scattering to constrain the number and size of additional multiplets, which will in turn limit the total number of possible models. In AC scenarios, the allowed multiplets are those of the generalized GM models~\cite{GGM}: AC triplet, AC quartet, AC pentet, or AC sextet. These models have a field content that can be written as $(N,N)$ of $SU(2)_{L} \times SU(2)_{R}$, and under the EW group $SU(2)_{L} \times U(1)_{Y}$ has the following particle decomposition: AC triplets have the representations (1,2) and (1,0). AC quartet have the representations (3/2,3) and (3/2,1). AC pentet have the representations (2,4), (2,2) and (2,0). AC sextet have the representations (5/2,5), (5/2,3) and (5/2,1).
We can also have different combinations of those models. In addition, we explore the possibility of introducing small custodial violation on the vevs which is constrained by the $\rho$ parameter. As we will discuss, those modifications are usually small, since the main source of custodial violation needs to come from the potential.

Following~\cite{GGM}, we construct a perturbative unitary bound by computing the scattering matrix in the scalar sector. The largest eigenvalue of the scattering matrix for a single complex scalar multiplet with size $n$ is given by:
\begin{align}
a_{0}(T) = \frac{g^{2}}{16 \pi } \frac{\sqrt{n}(n^{2}-1)}{2\sqrt{3}} \, .
\end{align}
For a real multiplet, the eigenvalue needs to be divided by $\sqrt{2}$. We impose the perturbative unitarity constraint $|\Re a_{0}| < 1/2$.  In the case with more than one multiplet, the largest eigenvalue of the overall scattering matrix is found by adding the eigenvalues for each multiplet in quadrature. Using this expression and considering the case with only one doublet, we have 4487 possible combinations of scalar multiplets that preserve perturbative unitarity. From these combinations, we can only have at most one AC sextet, four AC pentets, 23 AC quartets, or 145 AC triplets.  For this work, we study the following cases: AC triplet, AC quartet, AC pentet, AC sextet, AC pentet + AC sextet, and two AC pentet + AC sextet. Additionally, we also explore the case with general vevs for each of these models.

Each different multiplet has a coupling modifier for the vector bosons $V=W, \, Z$  that can be seen in Appendix~\ref{ap:kappas}. After the diagonalization to the mass basis, the Higgs will have a coupling modifier of the form:
\begin{align}
\kappa_{V}^{h} = R_{1}\kappa_{V}^{\text{doublet}} + R_{2}\kappa_{V}^{\text{multiplet 1 }} + ... +  R_{n} \kappa_{V}^{\text{multiplet (n-1) }}
\, .
\end{align}
The coupling modifiers are a function of the vacuum expectation value of the given multiplet, while the diagonalization vector $\vec{R}$ depends also on the specific potential. As noted already, we will assume that we have enough freedom on the potential such that we can treat $\vec{R}$ as a random unit vector that is independent of the vevs. Because of this assumption, we only need to generate a random vector and the vevs. Then, we check if it is possible to generate a negative $\lambda_{WZ}$, and also how the other observables behave in such a region.

Besides these models, there could be a situation with a large number of multiplets, each with a small vev, but with the total contribution to $\kappa$'s being order one. To study these possibilities, we pick the extremal cases of 145 AC triplets, 23 AC quartets, and 4 pentets. In these models we assume that the vevs are equal, but an arbitrary fraction of the total vev. In this case, the generation of the parameter space is challenging because of the number of free parameters. However, because we are interested only on the most extremal case, where the couplings can be as close as possible to the standard model, we can use the Cauchy–Schwarz inequality to re-write the $\kappa$'s in terms of only 3 random variables for the first two cases or 4 random variables for the 4 pentets case. In order to illustrate this, let us work the situation with $N$ copies of two electroweak multiplets\footnote{This encompasses the situation of 145 AC triplets and 23 AC quartets. The generalization for the 4 pentets is straighforward, by adding one additional variable $\tilde{R}_{4}$}:
 \begin{align} \nonumber
\kappa_{V}^{h} &= R_{1}\kappa_{V}^{\text{doublet}}  + \left( R_{2} \kappa_{V}^{\text{multiplet 1}}+R_{3} \kappa_{V}^{\text{multiplet 2}} \right) + \left( R_{4} \kappa_{V}^{\text{multiplet 1}}+R_{5} \kappa_{V}^{\text{multiplet 2}} \right) + \ldots = \, \\ \nonumber
 &=  R_{1}\kappa_{V}^{\text{doublet}} + \left(R_{2} + R_{4} + \ldots \right) \kappa_{V}^{\text{multiplet 1}} + \left( R_{3} + R_{5} + \ldots \right) \kappa_{V}^{\text{multiplet 2}} =  \,  \\
 &=  R_{1}\kappa_{V}^{\text{doublet}} + \tilde{R}_{2} \kappa_{V}^{\text{multiplet 1}} + \tilde{R}_{3} \kappa_{V}^{\text{multiplet 2}}
 \label{eq:kappa_many}
\end{align}
The inequality relations that can be constructed to bound $\tilde{R}_{2}$ and $\tilde{R}_{3}$ are the following:
\begin{align}
R_{1}^{2}&+\frac{\tilde{R}_{2}^{2}}{N} + \frac{\tilde{R}_{3}^{2}}{N} \leq 1 \, ,  \\
\abs{\tilde{R}_{2}} &\leq \sqrt{N}  \, , \, \abs{\tilde{R}_{3}} \leq \sqrt{N} \, .
\end{align}
Using these relations we can simplify the parameter space and also obtain the maximal contributions for the observables without having to resolve the degeneracy.

Now that the framework is set we can apply the Higgs data to the specific cases and try to understand the current state of such models.

\section{Experimental bounds on negative $\lambda_{WZ}$ for extended scalar sectors} \label{sec:EXP}

The experimental values that we use for this analysis are from ATLAS Higgs combination~\cite{ATLASlwz}. Our detailed statistical procedure, as well as an analysis of the CMS combination~\cite{CMSlwz} data, are given in Appendix~\ref{ap:fit}. Our analysis will use the fits for the ratios of coupling modifiers $\lambda_{WZ}$, $\lambda_{fZ}$ and $\kappa_{fZ}$ defined in Eqs.~\eqref{eq:lamwz}, \eqref{eq:lamfz}, and~\eqref{eq:kapfz}, respectively.
We fix $\lambda_{WZ}$ to be negative and conservatively allow it to be within the 5$\sigma$ allowed region:
\begin{equation}
-1.44 \leq \lambda_{WZ} \leq -0.69 .
\end{equation}
We can then see how the other ratios of coupling modifiers $\abs{\lambda_{fZ}}$ and $\abs{\kappa_{fZ}}$ behave given this constraint.  We could be more stringent and bound $\lambda_{WZ}$ to be inside the $3\sigma$ region and include the correlation in this parameter. As we will see, all models are excluded with this conservative bound, thus they will also be in the stringent one.
Note that we are ignoring the contributions of $\kappa_\gamma$ and $\kappa_{Z\gamma}$ to $\kappa_h$ as explained in footnote~\ref{floop}.

At this point, we should reinforce the reason that we are using the ratio of coupling modifiers, instead of the $\kappa$'s directly. In these models, there is the contribution of additional particles inside the processes $h\rightarrow \gamma \gamma$, $h \rightarrow Z \gamma$, and potential new Higgs decays. Although direct and indirect experimental constraints on the Higgs boson width exist, they are usually model dependent. Since $\Gamma_{h}$ is not experimentally constrained in a model-independent way, only ratios of coupling strengths can be measured in the most generic parametrisation considered in the $\kappa$ framework.
 This is important because the negative $\kappa_{V}$ is disfavoured in every other fit \cite{ATLASCMSCOMB}, especially because cancellations in the di-photon decay cannot be counteracted by new particles. The ATLAS analysis ignores the negative $\kappa_{V}$ region because of this, but ends up also ignoring this possibility in fits to the ratios of coupling modifiers.

\begin{figure*}[h!]
     \resizebox{0.49\linewidth}{!}{\includegraphics{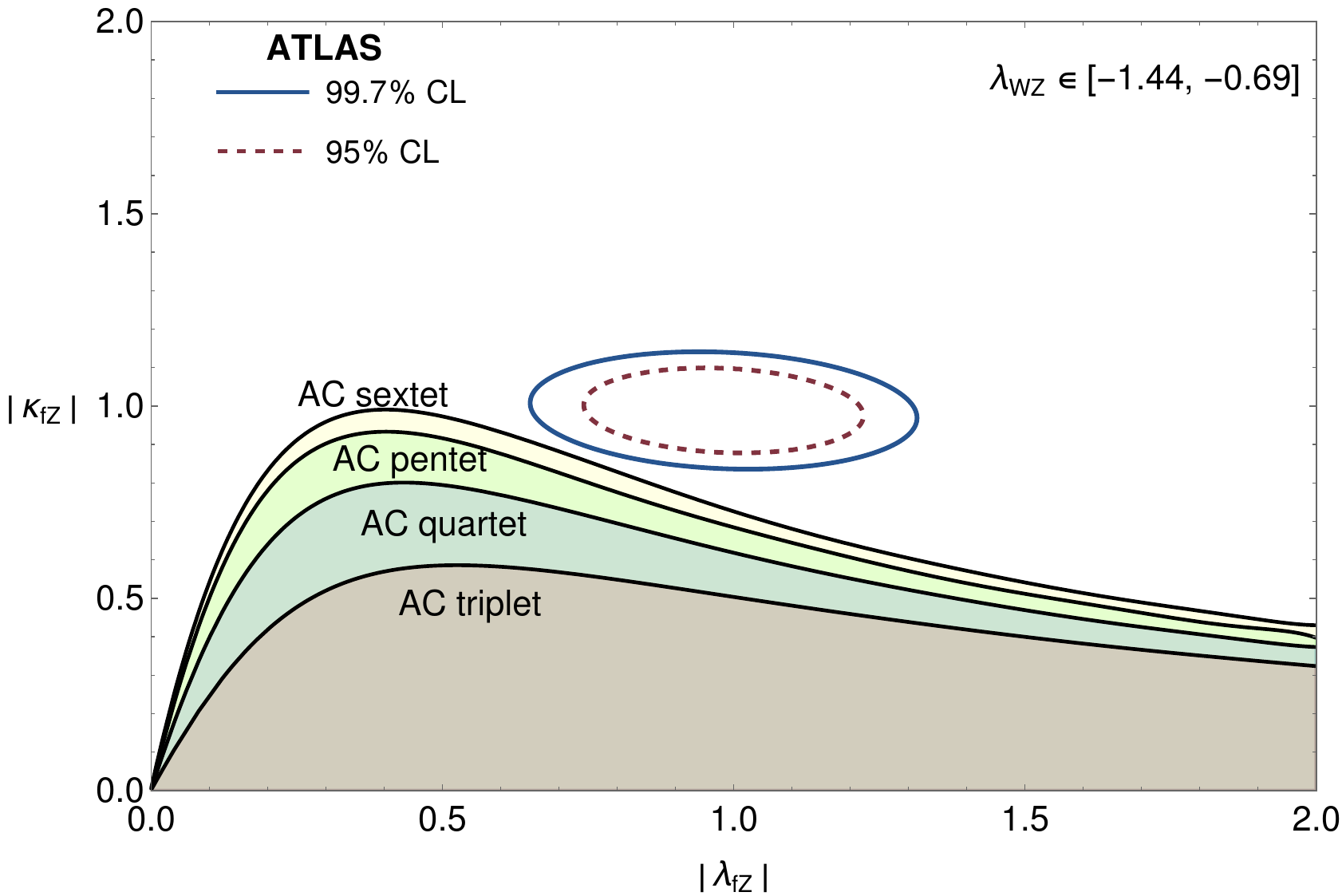}}
     \resizebox{0.49\linewidth}{!}{\includegraphics{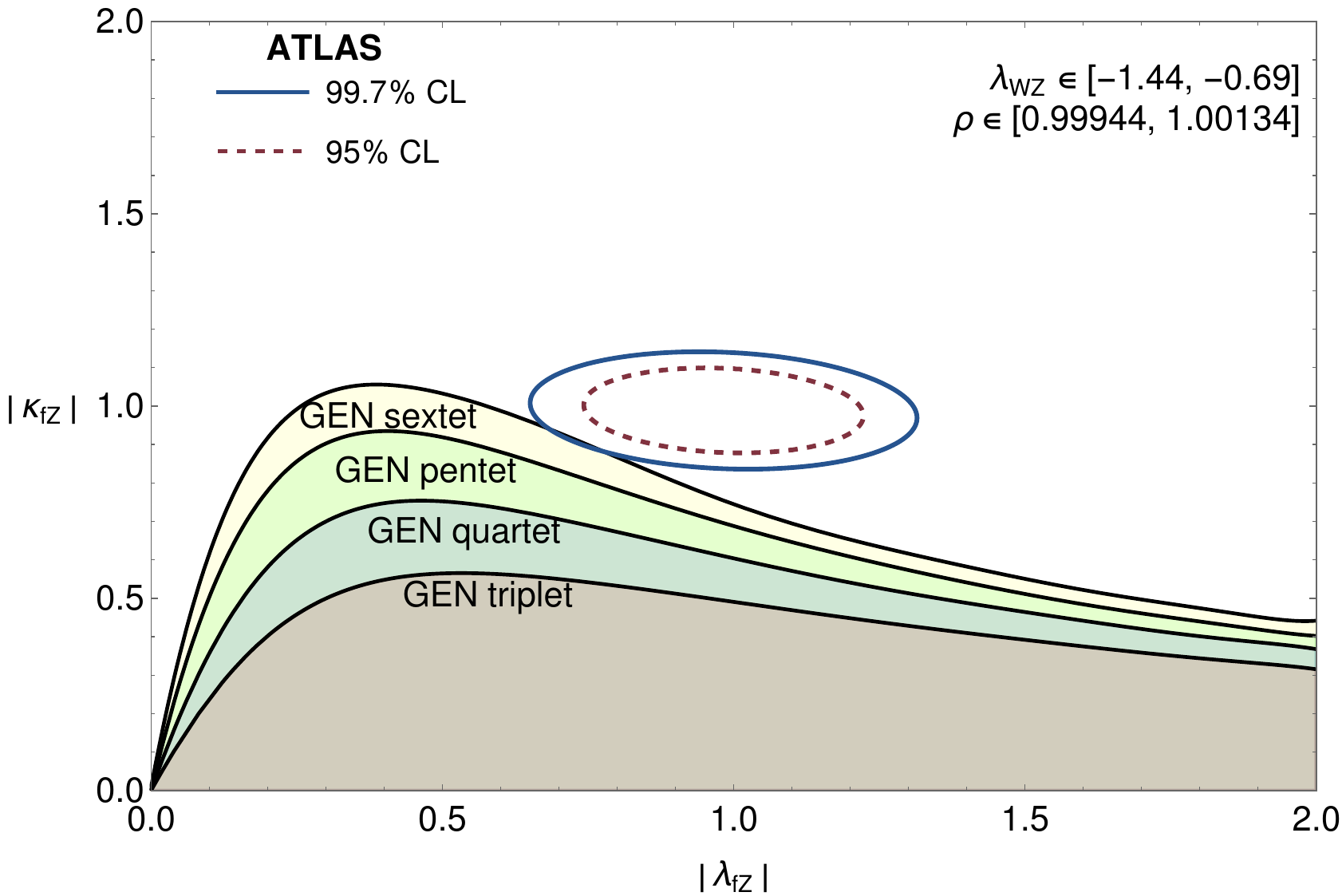}}
     \caption{ \label{fig:ONEACGEN} Accessible parameter space in $|\kappa_{fZ}|$ vs.~$|\lambda_{fZ}|$ plane for models with a single AC (general) multiplet on the left (right). We also show the $2$ and $3\sigma$ allowed regions from the ATLAS combination~\cite{ATLASlwz}.}
\end{figure*}

In the left panel of Figure~\ref{fig:ONEACGEN}, we give the accessible regions in the $|\lambda_{fZ}|$ and $|\kappa_{fZ}|$ plane for different models with a single AC multiplet. These curves were generated with a parameter scan, and the details are given in Appendix~\ref{ap:fitSCAN}. We then compare the allowed region to the ATLAS fit and see that these models are all excluded with greater than 99.7\% CL.  It is interesting to note that larger multiplets can come closer to the SM value of the couplings. This suggests that if we could go to arbitrarily large multiplets, then we could find a model that would still be allowed by the Higgs precision measurements. However, these models are excluded by perturbative unitarity, or at least, we cannot trust the perturbation theory. This shows that, at least for perturbative theories, all models with one AC multiplet are excluded in the negative $\lambda_{WZ}$ region.

On the right panel of Figure~\ref{fig:ONEACGEN}, we also include the possibility of custodial violation on the vacuum configuration. Such models are bounded by the $\rho$ parameter, and we again take a conservative $5\sigma$ allowed region for $\rho$:
\begin{equation}
0.99944 \leq \rho \leq 1.00134 .
\end{equation}
We use only the tree-level contributions to $\rho$ given in Eq.~\eqref{eq:rho-triplet} and its generalization to larger multiplets. Model dependant one-loop contributions could potentially be of similar size to the experimental precision on the $\rho$ parameter.
Therefore, in order to provide a model-independent analysis, we use the loose $5 \sigma$ relation and we can expect that the one-loop contributions will not significantly modify the allowed values. Loop contributions to $\rho$, and more generally bounds from precision electroweak and flavour physics, are model dependent and require knowing the full scalar potential and/or the particle spectrum of a particular model. These bounds can be computed, for example as in~\cite{indGM}, but computing them for all possible models would be technically challenging.

Comparing the two panels in Figure~\ref{fig:ONEACGEN} , we see that custodial violation on the potential is not important, it only slightly changes the curves, but does not change the overall picture. This occurs because the $\rho$ parameter is a strong constraint, even at $5\sigma$. The small change does push some regions of the GEN sextet model inside of the $3\sigma$ ellipse; this model is now excluded at $99.5\%$ CL, still well more than $2\sigma$.

\begin{figure*}[b!]
     \resizebox{0.49\linewidth}{!}{\includegraphics{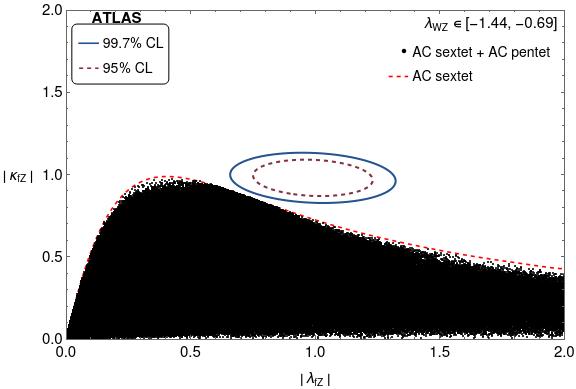}}
     \resizebox{0.49\linewidth}{!}{\includegraphics{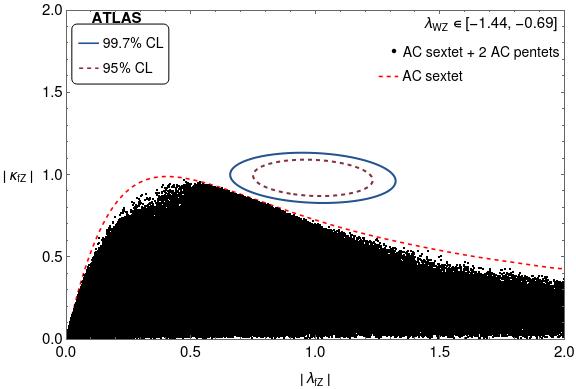}}
          \resizebox{0.49\linewidth}{!}{\includegraphics{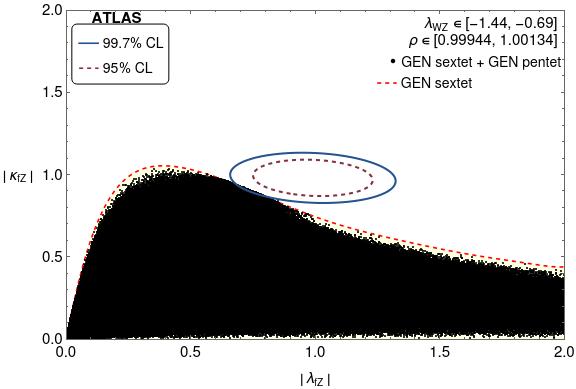}}
     \caption{ \label{fig:MULTACGEN} Scatter plots for the cases of AC sextet + AC pentet (left), AC sextet + 2 AC pentets (right) and general vevs for sextet + pentet (bottom). In all cases the expected behaviour of being equal or worse than the situation with only the largest multiplet is highlighted. }
\end{figure*}

To give a general picture of other possible scenarios, we can include the situations with more than one AC multiplet. The overall picture does not change much since the main contribution still comes from the largest multiplet. We can see in Figure~\ref{fig:MULTACGEN} that we are still away from the $3\sigma$ region and there is little difference between the case with only one AC sextet with the case where we add an AC pentet to the mix. The extension for the general vevs also shows the same trend as before and does not differ much from the custodial vacuum case. Note that in order to explore the boundary in Figure~\ref{fig:MULTACGEN} from a high dimensional parameter space scan, besides a random scan, {\tt MultiNest}~\cite{Feroz:2007kg,Feroz:2008xx,Feroz:2013hea} is used to scan the parameter which exploits nested sampling and automatically generates points close to the experiments contours.

It is worth pointing out one important thing about the behaviour of the $\kappa$'s in the case with more than one multiplet. One would expect given that models with more multiplets have more free parameters, that those models have more freedom to generate large contributions to the coupling modifiers. However, it is possible to see in Figure~\ref{fig:MULTACGEN} that adding more multiplets does not bring the coupling modifiers closer to the Standard Model values or raise the location of the bounding curve. The contribution is always equal to or smaller than the case with only the largest multiplet, which in this case is the sextet. This behaviour can be traced to the constraint on the total EW vev. The more multiplets share the EW symmetry breaking, the smaller is the individual contribution. In turn, the lower possible values of vevs constrains the maximum contribution for the $\kappa$'s and thus suppresses the coupling modifiers. This effect is greater than the possibility of accidental cancellations between multiplets to enhance the couplings. Then, adding more multiplets makes it harder to generate a negative $\lambda_{WZ}$, not easier.

\begin{figure*}[b!]
     \resizebox{0.49\linewidth}{!}{\includegraphics{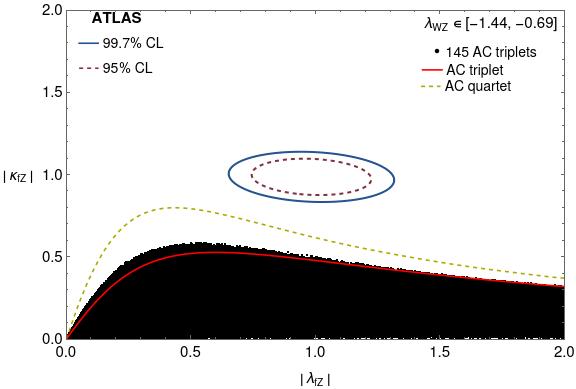}}
     \resizebox{0.49\linewidth}{!}{\includegraphics{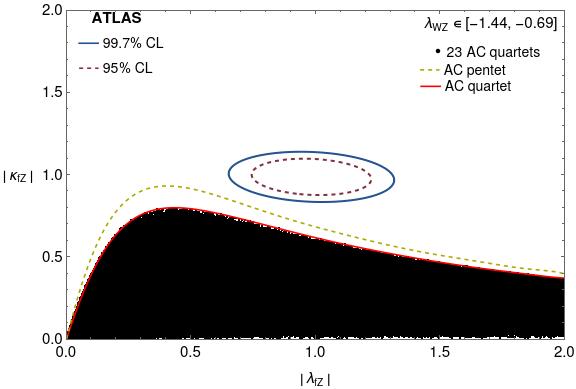}}
          \resizebox{0.49\linewidth}{!}{\includegraphics{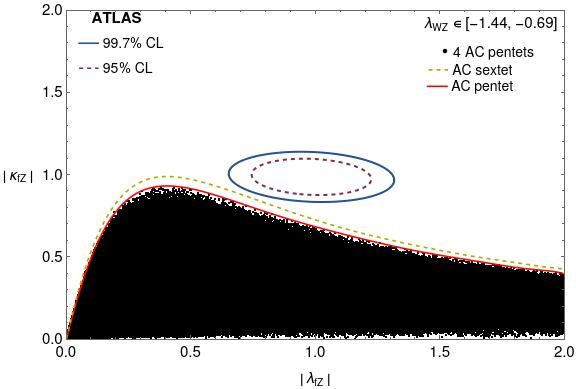}}
     \caption{ \label{fig:MULTMULT} Scatter plots for the cases of 145 AC triplets (left), 23 AC quartets (right) and 4 AC pentets (bottom). The expectation values for each case is assumed to be equal and a random fraction of the total vev.  }
\end{figure*}

To highlight this behavior, we consider the possibility of a large number of multiplets with equal vevs as described in detail in Eq.~\eqref{eq:kappa_many}. One could imagine that the contribution of each multiplet is small, but they add to make a large modification to the coupling modifiers. This turns out not to be the case as shown in Figure~\ref{fig:MULTMULT}, where the scenarios with multiple multiplets are always further from the allowed region than model with the next largest multiplet. Because of that, we can conjecture that all weakly coupling extensions of the Higgs sector have the negative coupling region excluded, independent of the number of multiplets.

\section{Conclusion} \label{sec:CONC}

In this work, we studied the current status of negative coupling modifiers in extended Higgs sectors, with the focus on the observable $\lambda_{WZ}$ which measures the amount of custodial violation in the model. The experimental data has bounded $|\lambda_{WZ}| \approx 1$, but there is currently very little information on its sign.  We present the class of extended scalar sectors (AC multiplets) that have the best chance of generating $\lambda_{WZ}\approx -1$, while avoiding $\rho$ parameter constraints. We analyze the simplest case of AC triplets, and then we show how to generalize the procedure to different multiplets.

The possibility of exploring this wide range of models lies in the fact that the coupling modifiers, in the end, depend only on the diagonalization matrix and the vevs. Thus we assume that the potential has enough parameters such that we can treat the eigenvector as a random unit vector that is uncorrelated with the vevs. This approach is more general than doing the individual potential scans, and it is useful when one is looking for the exclusion of parameter regions. With this tool at hand, we explore different models that have a custodial vacuum.

Our analysis shows that all the models with one or more AC multiplets studied here are excluded by the ATLAS~\cite{ATLASlwz} results at $99.5\%~\text{CL}$. in Appendix~\ref{ap:fit} we also compare to the CMS data~\cite{CMSlwz} and show the exclusion is larger than $95\%$ CL. Making a more precise statement would require the correlation of the coupling modifiers. This result stays almost the same even when we allow for custodial breaking vacua on these models. This was expected, since the $\rho$ parameter bound is very strong, even at using a conservative $5\sigma$ constraint.

In the analysis with multiple AC multiplets, we can see the effect of suppression on the parameter space, moving away from the experimental central value. We can understand that if we add more AC sectors together, the overall behaviour is still dominated by the largest multiplet. Because of that, we can see that for all the cases explored here, these models are excluded at $99.5 \%$ CL by the ATLAS data. This exclusion will get stronger with future HL-LHC data. We can also conjecture that all weakly coupled extensions of the Higgs sector have the region with $\lambda_{WZ} < 0$ excluded, independent of the number of multiplets.

What does this mean for negative  $\lambda_{WZ}$? In any weakly coupled model, the only way to acquire such values are with the use of extended scalar sectors. There could be non-perturbative effects that achieve the same feature, but this is not currently known. We can then say that this region of parameter space is heavily disfavoured for any weakly coupled extended scalar sector. The precision that we acquire in the Higgs sector now is enough to detect this accidental cancellation and has a powerful consequence for what can be beyond the SM . This removes another potentially large custodial violation source of the new physics, showing that custodial violation is likely a good symmetry of nature.

In contrast, if the measured best fit value for the CMS fit remains negative with more data and different experiments confirm this, we would not be able to describe the new physics using the current methods. This would indicate the necessity of expanding the current understanding of extended scalar sectors in the non-perturbative domain. It may be that new physics is hiding in plain sight, after all, only future experiments can tell.

\section*{Acknowledgments}
\noindent We would like to thank Heather Logan for helpful discussions. C.H.dL.~and D.S.~are supported in part
by the Natural Sciences and Engineering Research Council of Canada (NSERC). Y.W. thanks the U.S.~Department of Energy for the financial support, under grant number DE-SC 0016013. Part of the computation for this project was performed at the High Performance Computing Center at Oklahoma State University, supported in part through the National Science Foundation grant OAC-1531128.

\appendix

\section{Statistical combination of ATLAS and CMS ratios of coupling modifiers.}\label{ap:fit}

To know if a model is excluded, we need the experimental measurements for the observables. Because the models described in this work have extended representations, we need to be careful with the existence of new particles inside the loops for Higgs to di-photon decay. Additionally, we assume there are no new fields that carry colour so the modification for the di-gluon decay occurs only through the fermion coupling, this means that for these extensions we have $\kappa_{g} = \kappa_{f}$. The important observables that we use are the following:
\begin{align}
\kappa_{fZ} = \frac{\kappa_{f}\kappa_{Z}}{\kappa_{h}} \, , \\
\lambda_{WZ} = \frac{\kappa_{W}}{\kappa_{Z}} \, , \\
\lambda_{fZ} = \frac{\kappa_{f}}{\kappa_{Z}} \, ,
\end{align}
with $\kappa_h$ defined in Eq.~\eqref{eq:kap-h}. We do not consider the modification from $\kappa_{\gamma}$ since this is model dependent and will be bound by other observables

 As mentioned before, the CMS measurement \cite{CMSlwz} indicates a negative central value for $\lambda_{WZ}$. However, in their work, there is no information on the correlation of the other ratios of coupling modifiers. This is in contrast with the ATLAS results \cite{ATLASlwz}. Therefore we mainly use the ATLAS result, and we assume that it is measuring only the absolute value of the coupling modifiers. The observables that we used are the following: $\kappa_{gZ}, \, \lambda_{Zg}, \, \lambda_{WZ}, \, \lambda_{\tau Z}, \, \lambda_{b Z}$. We combine $1/\lambda_{Zg}, \, \lambda_{\tau Z}$, and $ \lambda_{b Z}$ into $\lambda_{fZ}$ with a least squares fit. We do not include the correlation for $\lambda_{WZ}$ to preserve the shape of the 2D plane. The inclusion of this additional correlation would only make the results discussed in this work stronger since they shrink the allowed parameter space for $\lambda_{WZ}$. From the combination we have the following central value and covariance matrix from the ATLAS data:
\begin{align}
(\lambda_{fZ},\kappa_{fZ} ) &= (0.99, 0.98) \, , \\
\text{COV} &= \begin{pmatrix}
0.0093 & -0.00054 \\
-0.00054 & 0.0020
\end{pmatrix} \, , \\
\lambda_{WZ} &= 1.04_{-0.07}^{+ 0.08} \, .
\end{align}

\begin{figure*}[h!]
     \resizebox{0.49\linewidth}{!}{\includegraphics{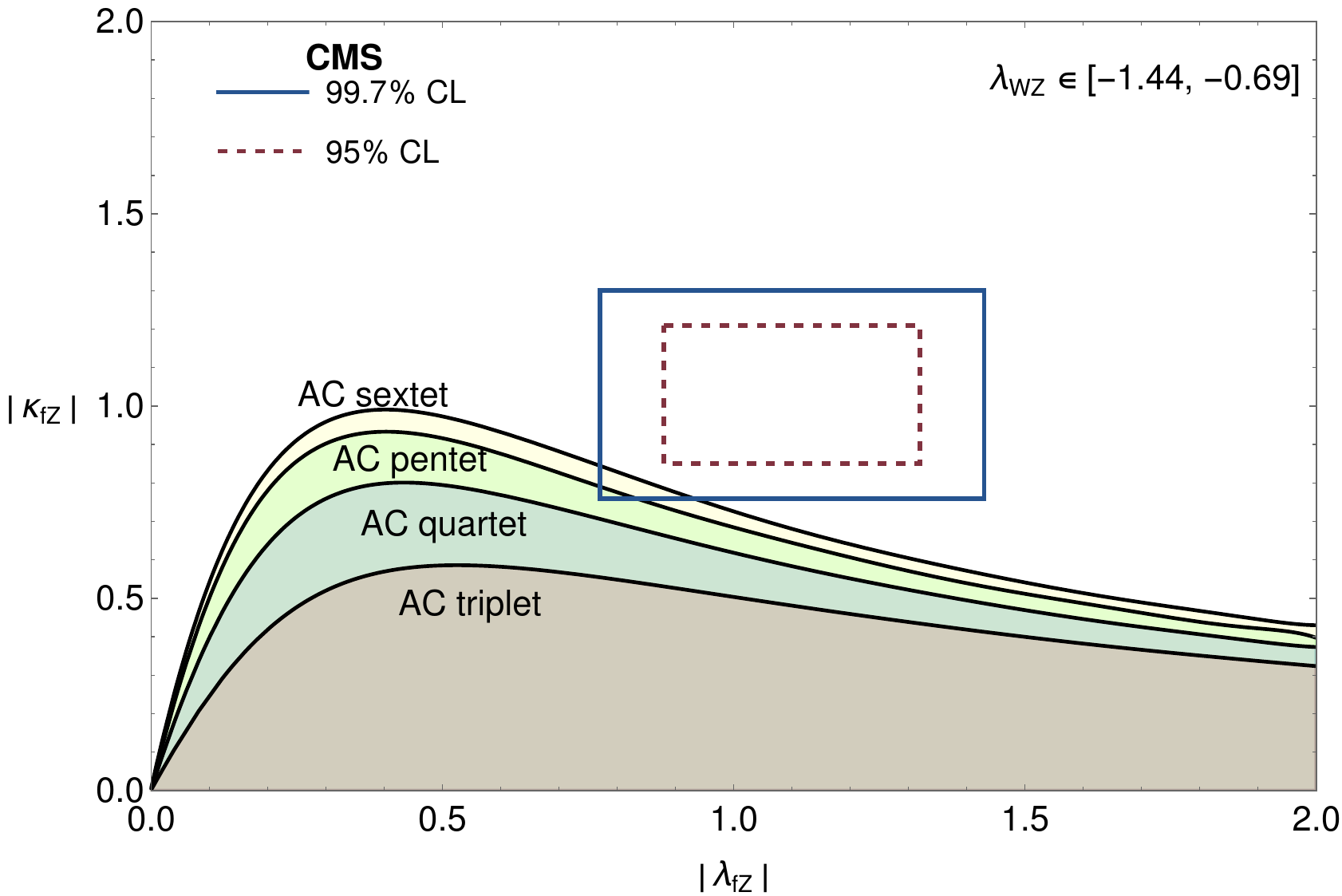}}
     \resizebox{0.49\linewidth}{!}{\includegraphics{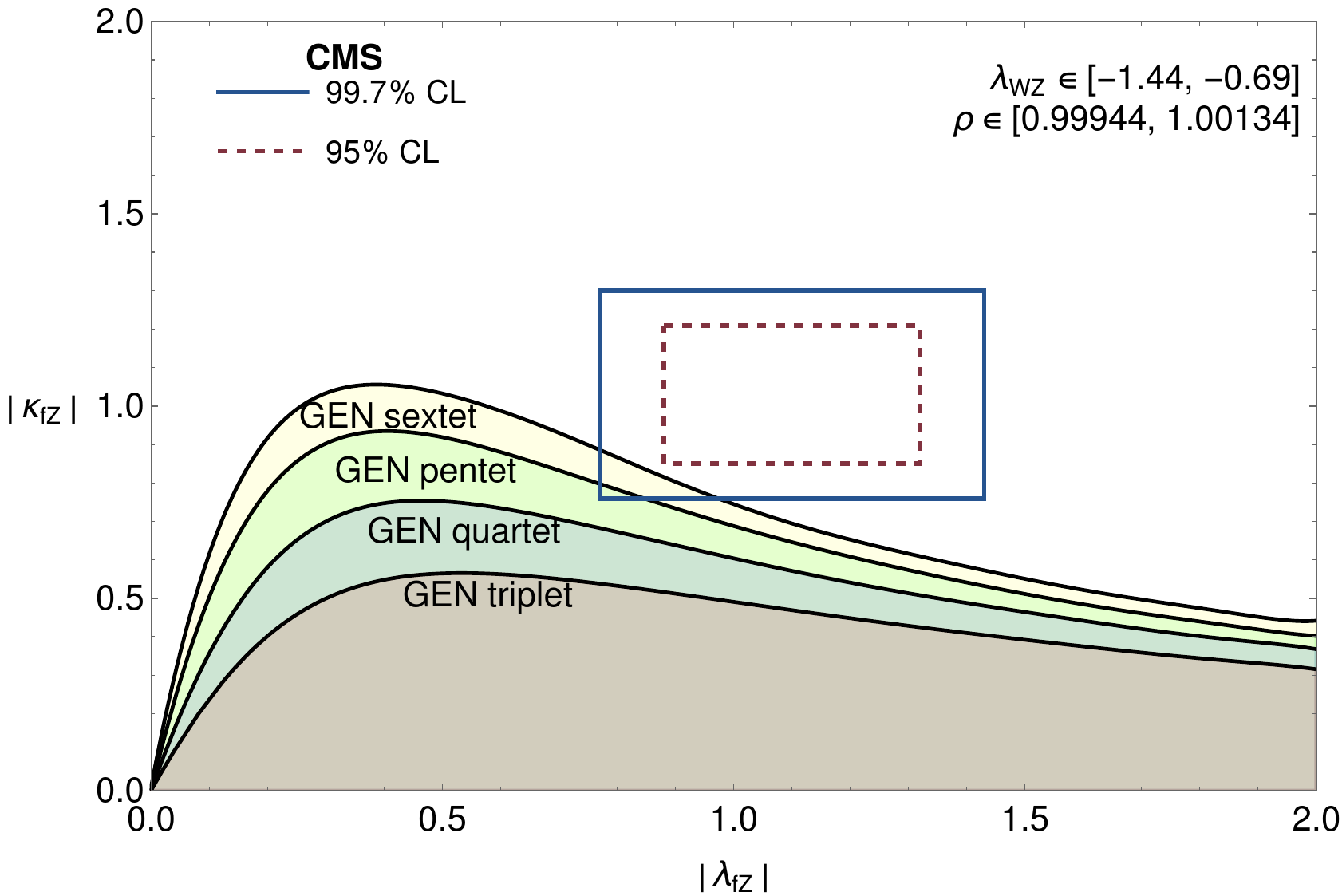}}
     \caption{ \label{fig:ONEACGENCMS} Relation between  $|\lambda_{fZ}|$ and $|\kappa_{fZ}|$ for  $\lambda_{WZ}$ negative and inside the $3\sigma$ region of CMS values assuming no correlations for models with one AC multiplet (left) or general vevs (right).}
\end{figure*}

The fit for CMS uses the same observables.  The measured values from the CMS fit are:
\begin{align}
\kappa_{fZ} &= 1.03 \pm 0.09 \, , \\
\lambda_{fZ} &= 1.10 \pm 0.11 \, , \\
\lambda_{WZ} &= -1.13^{+0.10}_{-0.11} \, .
\end{align}
In our analysis, the standard deviation of $\kappa_{fZ}$ also plays an important role, and as one can notice, the precision from ATLAS is better than of CMS. Analyzing the CMS data without correlations, we do have allowed points inside the $3\sigma$ region, but they are excluded at 95\% CL. The results for the uncorrelated CMS measurement seen in Figure~\ref{fig:ONEACGENCMS}. Including correlation will  make this result stronger; if the correlation is be similar to ATLAS, the allowed values would be outside $99.7\%$ CL region.

\section{Group theory basis}\label{ac:group}

In this Appendix we give the basis for the generators that we use for different $SU(2)$ multiplets.
The doublet generators are:
\begin{align}
\tau_{1} = \left(
\begin{array}{cc}
 0 & \frac{1}{2} \\
 \frac{1}{2} & 0 \\
\end{array}
\right) \, , \, \tau_{2} = \left(
\begin{array}{cc}
 0 & -\frac{i}{2} \\
 \frac{i}{2} & 0 \\
\end{array}
\right) \, , \, \tau_{3} = \left(
\begin{array}{cc}
 \frac{1}{2} & 0 \\
 0 & -\frac{1}{2} \\
\end{array}
\right)  \, .
\end{align}
The triplet generators are:
\begin{align}
t_{1} = \left(
\begin{array}{ccc}
 0 & \frac{1}{\sqrt{2}} & 0 \\
 \frac{1}{\sqrt{2}} & 0 & \frac{1}{\sqrt{2}} \\
 0 & \frac{1}{\sqrt{2}} & 0 \\
\end{array}
\right) \, , t_{2} = \left(
\begin{array}{ccc}
 0 & -\frac{i}{\sqrt{2}} & 0 \\
 \frac{i}{\sqrt{2}} & 0 & -\frac{i}{\sqrt{2}} \\
 0 & \frac{i}{\sqrt{2}} & 0 \\
\end{array}
\right) \, , \, t_{3} = \left(
\begin{array}{ccc}
 1 & 0 & 0 \\
 0 & 0 & 0 \\
 0 & 0 & -1 \\
\end{array}
\right) \, .
\end{align}
The quartet generators are:
\begin{align}
t_{1} = \left(
\begin{array}{cccc}
 0 & \frac{\sqrt{3}}{2} & 0 & 0 \\
 \frac{\sqrt{3}}{2} & 0 & 1 & 0 \\
 0 & 1 & 0 & \frac{\sqrt{3}}{2} \\
 0 & 0 & \frac{\sqrt{3}}{2} & 0 \\
\end{array}
\right) \, , \, t_{2} = \left(
\begin{array}{cccc}
 0 & -\frac{i \sqrt{3}}{2} & 0 & 0 \\
 \frac{i \sqrt{3}}{2} & 0 & -i & 0 \\
 0 & i & 0 & -\frac{i \sqrt{3}}{2} \\
 0 & 0 & \frac{i \sqrt{3}}{2} & 0 \\
\end{array}
\right) \, , \, t_{3} = \left(
\begin{array}{cccc}
 \frac{3}{2} & 0 & 0 & 0 \\
 0 & \frac{1}{2} & 0 & 0 \\
 0 & 0 & -\frac{1}{2} & 0 \\
 0 & 0 & 0 & -\frac{3}{2} \\
\end{array}
\right) \, .
\end{align}
The pentet generators are:
\begin{align}
t_{1} = \left(
\begin{array}{ccccc}
 0 & 1 & 0 & 0 & 0 \\
 1 & 0 & \sqrt{\frac{3}{2}} & 0 & 0 \\
 0 & \sqrt{\frac{3}{2}} & 0 & \sqrt{\frac{3}{2}} & 0 \\
 0 & 0 & \sqrt{\frac{3}{2}} & 0 & 1 \\
 0 & 0 & 0 & 1 & 0 \\
\end{array}
\right) \, , \, t_{2}= \left(
\begin{array}{ccccc}
 0 & -i & 0 & 0 & 0 \\
 i & 0 & -i \sqrt{\frac{3}{2}} & 0 & 0 \\
 0 & i \sqrt{\frac{3}{2}} & 0 & -i \sqrt{\frac{3}{2}} & 0 \\
 0 & 0 & i \sqrt{\frac{3}{2}} & 0 & -i \\
 0 & 0 & 0 & i & 0 \\
\end{array}
\right) \, , \, t_{3}= \left(
\begin{array}{ccccc}
 2 & 0 & 0 & 0 & 0 \\
 0 & 1 & 0 & 0 & 0 \\
 0 & 0 & 0 & 0 & 0 \\
 0 & 0 & 0 & -1 & 0 \\
 0 & 0 & 0 & 0 & -2 \\
\end{array}
\right) \, .
\end{align}
The sextet generators are:
\begin{align}
t_{1} &= \left(
\begin{array}{cccccc}
 0 & \frac{\sqrt{5}}{2} & 0 & 0 & 0 & 0 \\
 \frac{\sqrt{5}}{2} & 0 & \sqrt{2} & 0 & 0 & 0 \\
 0 & \sqrt{2} & 0 & \frac{3}{2} & 0 & 0 \\
 0 & 0 & \frac{3}{2} & 0 & \sqrt{2} & 0 \\
 0 & 0 & 0 & \sqrt{2} & 0 & \frac{\sqrt{5}}{2} \\
 0 & 0 & 0 & 0 & \frac{\sqrt{5}}{2} & 0 \\
\end{array}
\right) \, , \, t_{2} = \left(
\begin{array}{cccccc}
 0 & -\frac{i \sqrt{5}}{2} & 0 & 0 & 0 & 0 \\
 \frac{i \sqrt{5}}{2} & 0 & -i \sqrt{2} & 0 & 0 & 0 \\
 0 & i \sqrt{2} & 0 & -\frac{3 i}{2} & 0 & 0 \\
 0 & 0 & \frac{3 i}{2} & 0 & -i \sqrt{2} & 0 \\
 0 & 0 & 0 & i \sqrt{2} & 0 & -\frac{i \sqrt{5}}{2} \\
 0 & 0 & 0 & 0 & \frac{i \sqrt{5}}{2} & 0 \\
\end{array}
\right) \, , \\
t_{3} &= \left(
\begin{array}{cccccc}
 \frac{5}{2} & 0 & 0 & 0 & 0 & 0 \\
 0 & \frac{3}{2} & 0 & 0 & 0 & 0 \\
 0 & 0 & \frac{1}{2} & 0 & 0 & 0 \\
 0 & 0 & 0 & -\frac{1}{2} & 0 & 0 \\
 0 & 0 & 0 & 0 & -\frac{3}{2} & 0 \\
 0 & 0 & 0 & 0 & 0 & -\frac{5}{2} \\
\end{array}
\right) \, .
\end{align}

\section{Coupling modifiers for different multiplets}\label{ap:kappas}

To know the total coupling modifier for the Higgs, we need the contributions from the different gauge multiplets. Here we work out the different states that can have a custodial preserving vacuum. In this notation, the custodial limit is the one where all the vevs for each set of fields are equal. The general notation for the neutral component vev is:
\begin{align}
<\phi_{(I,Y)}> = \nu_{(I,Y)}
\end{align}
The only difference is the doublet vev which we introduce an factor of $1/ \sqrt{2}$. The coupling modifiers are defined as:
\begin{align}
\kappa_{i}^{(I,Y)} =\frac{ g_{i}^{(I,Y)}}{ g_{i}^{SM}} \, .
\end{align}

 First, we have the standard doublet with quantum numbers $SU(2)_{L} \times U(1)_{Y} = (1/2,1)$ that generates the following contributions:
\begin{align}
\kappa_{f}^{(1/2,1)} =  \frac{\nu_{(1/2,1)} }{\nu} \, , \,  \kappa_{W}^{(1/2,1)} = \frac{\nu_{(1/2,1)} }{\nu} \, , \, \kappa_{Z}^{(1/2,1)} = \frac{\nu_{(1/2,1)} }{\nu}  \, ,
\end{align}
where $\nu$ is the total electroweak vev defined in Eq.~\eqref{eq:vGf}.
For the AC triplet, we have one field with $(1,2)$ and another with  $(1,0)$ quantum numbers, then the coupling modifiers are:
\begin{align}
\kappa_{f}^{(1,2)} &= 0 \, , \,  \kappa_{W}^{(1,2)} = \frac{2\sqrt{2} \nu_{(1,2)}}{\nu}  \, , \, \kappa_{Z}^{(1,2)} = \frac{4\sqrt{2}\nu_{(1,2)}}{\nu} \,  \, , \\
\kappa_{f}^{(1,0)} &= 0 \, , \,  \kappa_{W}^{(1,0)} = \frac{4\nu_{(1,2)}}{\nu}  \, , \, \kappa_{Z}^{(1,0)} = 0 \, .
\end{align}
The AC quartet has one field with $(3/2,3)$ and another with $(3/2,1)$ quantum numbers, and the coupling modifiers are:
\begin{align}
\kappa_{f}^{(3/2,3)} &= 0 \, , \,  \kappa_{W}^{(3/2,3)} = \frac{3\sqrt{2}\nu_{(3/2,3)}}{\nu}  \, ,\, \kappa_{Z}^{(3/2,3)} =\frac{9\sqrt{2}\nu_{(3/2,3)}}{\nu} \,  \, , \\
\kappa_{f}^{(3/2,1)} &=0  \, , \,  \kappa_{W}^{(3/2,1)} = \frac{7\sqrt{2}\nu_{(3/2,1)}}{\nu}  \, , \, \kappa_{Z}^{(3/2,1)} = \frac{\sqrt{2}\nu_{(3/2,1)}}{\nu}\, .
\end{align}
\begin{figure*}[t!]
     \resizebox{0.49\linewidth}{!}{\includegraphics{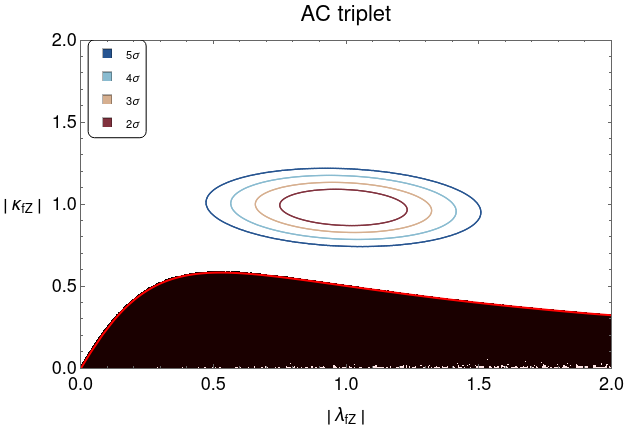}}
     \resizebox{0.49\linewidth}{!}{\includegraphics{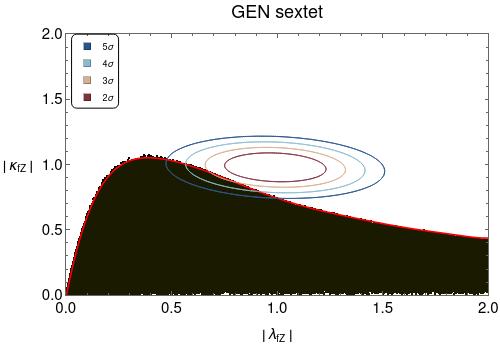}}
     \caption{ \label{fig:FITAPP} Relation between the scatter plot and the fit in the $|\lambda_{fZ}|$ and $|\kappa_{fZ}|$ plane for two specific models. As in all previous figures, $\lambda_{WZ}$ is negative and inside the $5\sigma$ region of the ATLAS combination. Unlike in previous figures, we also include the 4 and $5\sigma$ exclusion contours.}
\end{figure*}
For the AC pentet, we have $(2,4)$, $(2,2)$ and $(2,0)$ with the following coupling modifiers:
\begin{align}
\kappa_{f}^{(2,4)} &= 0 \, , \,  \kappa_{W}^{(2,4)} = \frac{4\sqrt{2}\nu_{(2,4)}}{\nu}  \, , \, \kappa_{Z}^{(2,4)} = \frac{16\sqrt{2}\nu_{(2,4)}}{\nu} \,  \, , \\
\kappa_{f}^{(2,2)} &= 0 \, , \,  \kappa_{W}^{(2,2)} = \frac{10\sqrt{2} \nu_{(2,2)}}{\nu}  \, , \, \kappa_{Z}^{(2,2)} = \frac{4\sqrt{2}\nu_{(2,2)}}{\nu} \, , \\
\kappa_{f}^{(2,0)} &= 0 \, , \,  \kappa_{W}^{(2,0)} = \frac{12\nu_{(2,0)}}{\nu} \, , \, \kappa_{Z}^{(2,0)} = 0 \, .
\end{align}
Finally for AC sextet, we have $(5/2,5)$, $(5/2,3)$ and $(5/2,1)$:
\begin{align}
\kappa_{f}^{(5/2,5)} &= 0 \, , \,  \kappa_{W}^{(5/2,5)} = \frac{5\sqrt{2}\nu_{(5/2,5)}}{\nu}  \, , \, \kappa_{Z}^{(5/2,5)} = \frac{25\sqrt{2}\nu_{(5/2,5)}}{\nu} \,  \, , \\
\kappa_{f}^{(5/2,3)} &= 0 \, , \,  \kappa_{W}^{(5/2,3)} = \frac{13\sqrt{2}\nu_{(5/2,3)}}{\nu} \, , \, \kappa_{Z}^{(5/2,3)} = \frac{9\sqrt{2}\nu_{(5/2,3)}}{\nu} \, , \\
\kappa_{f}^{(5/2,1)} &= 0 \, , \,  \kappa_{W}^{(5/2,1)} = \frac{17\sqrt{2}\nu_{(5/2,1)}}{\nu}  \, , \, \kappa_{Z}^{(5/2,1)} = \frac{\sqrt{2}\nu_{(5/2,1)}}{\nu}\, .
\end{align}

\section{Fit of the individual scans} \label{ap:fitSCAN}

In this appendix, we want to highlight the fitting method employed to generate the solid curves in Fig.~\ref{fig:ONEACGEN}. The procedure is the following:  we generate a random scan using the methods described in the paper. Then, from the random scan, we generate a scatter plot in the $|\kappa_{fZ}|$ vs $|\lambda_{fZ}|$ plane. To obtain the boundary of the scatter plot, for every point in the scan, we find the point that has the largest value of $|\kappa_{fZ}|$ among points with $|\lambda_{fZ}|$ similar to the original point.
This list of maximal values will then be a noisy approximation to the boundary. We can then fit a smooth curve to this list to obtain the solid curves shown in Fig.~\ref{fig:ONEACGEN}. Using this method, depending on the nature of the points, there could be a small number of points above the contour curve. However, since the exclusion is at $99.5\%$ CL, this method is still safe and makes it easier to compare the information from different models. The comparison between the scatter and the fit can be seen in Figure \ref{fig:FITAPP} for some specific models. The few points that lie above the boundary are still excluded with well more than 95\% confidence.

\newcommand{\fakesection}[1]{%
  \par\refstepcounter{section}
  \sectionmark{#1}
  \addcontentsline{toc}{section}{\protect\numberline{\thesection}#1}
}

\fakesection{E}\makeatletter\def\@currentlabel{Erratum}\makeatother
\label{errata}

\newpage

\newcommand{\secondtitle}{
    \begin{center}
        \textbf{\large Erratum: Status of negative coupling modifiers for extended Higgs sectors}\\[1em]
    \end{center}
}

\secondtitle

Upon revisiting this work, we have found that there was a critical mistake in the expression of the fermion coupling modifiers in Eq.~\eqref{eq:tag}. While the rest of the analysis logic was correct, the mistake in that equation, when propagated through the analysis, substantially changed the conclusions of the paper regarding the experimental status of negative $\lambda_{WZ}$. Upon correcting the error, the accidentally custodial (AC) triplets described in section II are still excluded in a model-independent way. The AC quartets can be excluded in a model-independent way by including the weaker bound from $b \rightarrow s \gamma$. Unfortunately, all other AC models have vev structures such that it is possible to make the absolute values of coupling modifiers close to the SM prediction, and thus impossible to exclude those additional models using the methodology of our work. This weakens our claims of complete exclusion of negative $\lambda_{WZ}$ to a softer version where only the minimal models are excluded. On the other hand, this means that there are still models that can have significant coupling modifiers hidden from current data.

To highlight the difference in the analysis we can start by writing the correct coupling modifiers for the fermion-scalar interaction:
\begin{align}
\kappa_{f}^{\phi} = \frac{\nu}{\nu_{\phi}} \, .
\end{align}
Unlike what is written in the manuscript, the correct formula implies $\kappa_f^\phi \geq 1$. This changes the discussion starting at Eq.~(2.21) for the AC triplets. Enforcing $\lambda_{WZ}=-1$:
\begin{align}
R_{3} = - \frac{\nu_{\chi}}{\nu_{\phi}} \left(3\sqrt{2}R_{1}+2R_{2} \right) \, .
\end{align}
Next, we use this relation into $\lambda_{fZ}$ and enforce that $\abs{\lambda_{fZ}}=1$:
\begin{align}
R_{2} = \frac{R_{1}}{2} \left(\sqrt{2} \mp \frac{\nu^{2}}{\nu_{\chi}\nu_{\phi}} \right) \, .
\end{align}
We then check the last coupling modifier $\kappa_{fZ}$:
\begin{align}
\kappa_{fZ} = \pm R_{1} \frac{\nu}{\nu_{\phi}} \, . 
\label{eq:fz}
\end{align}
The problem now lies in the fact that $\frac{\nu}{\nu_{\phi}} $ can be larger than one. This implies that Eq.~\eqref{eq:fz} can generate values close to the SM prediction of $\kappa_{fZ}=1$. We can verify if this is the case by enforcing this condition to be the SM prediction and checking the normalization of the vector $\vec{R}$. This condition reads:
\begin{align}
\frac{\nu^{2}}{4\nu_{\chi}^{2}} - \frac{5\nu_{\chi}^{2}}{\nu^{2}} &= 1\, ,  \,  \, \text{for} \, \, \lambda_{fZ} = \kappa_{fZ} = 1 \, , \,  \\
\frac{1}{2} + \frac{3 \nu^{2}}{16 \nu_{\chi}^{2}} - \frac{5\nu_{\chi}^{2}}{\nu^{2}} &= 1 \, ,  \,  \, \text{for} \, \, \lambda_{fZ} = \kappa_{fZ} = -1 \, .
\end{align}
In this case, it is clear that this parameter point is excluded, since this condition is only satisfied for values of $\nu_{\chi}$ higher than the upper bound of $\nu_{\chi} \leq \frac{\nu}{2\sqrt{2}}$. This demonstrates that the $
\lambda_{WZ} =-1 $ scenario with AC triplets are still disfavored.

The situation is different for the AC quartets. Even with the same degree of freedom from $\vec{R}$, the group theory factors change the excluded window. Doing the same analysis we find that the coupling modifiers can be equal to the SM prediction if we have the AC quartet vacuum expectation values (vevs):
\begin{align}
\nu_{\omega} &\approx 0.198\nu ,  \,  \, \text{for} \, \, \lambda_{fZ} = \kappa_{fZ} = 1 \, , \,  \\
\nu_{\omega} &\approx 0.193\nu ,  \,  \, \text{for} \, \, \lambda_{fZ} = \kappa_{fZ} = -1 \, .
\end{align}
This is inside the allowed range for the vev of $\nu_{\omega} \lesssim 0.223\nu$. This condition means that the SM limit of the coupling modifiers implies a sizeable contribution from the AC quartets to the EWSB. We can use, in this case, a  bound from $b\rightarrow s\gamma$~\cite{GGM} to exclude the higher values of $\nu_\omega$ in a model-independent way and thus making the wrong-sign SM limit disfavored in the AC quartets model. 

\begin{figure}[t!]
\centering
    \includegraphics[width=0.49 \textwidth]{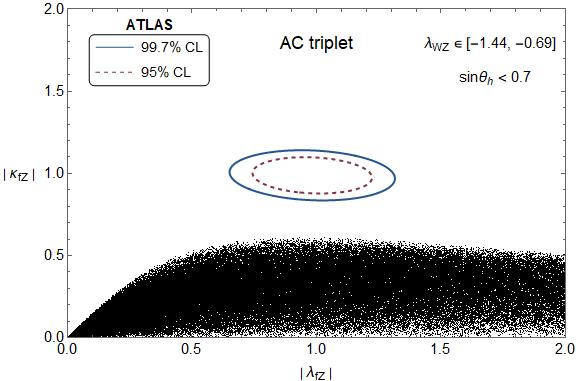}
    \includegraphics[width=0.49 \textwidth]{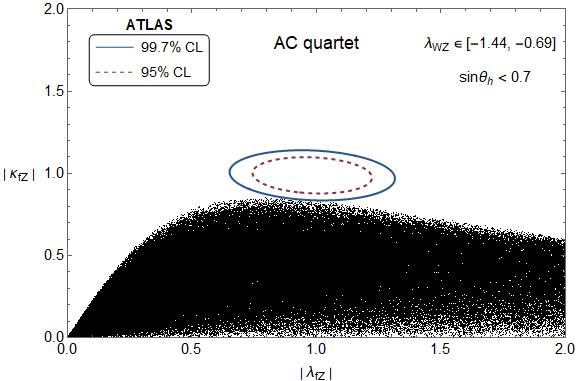}
    \includegraphics[width=0.49 \textwidth]{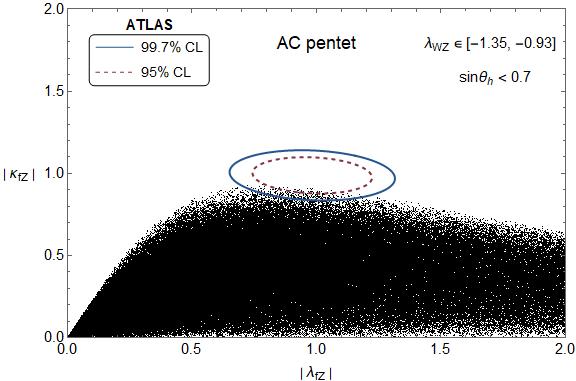} 
    \includegraphics[width=0.49 \textwidth]{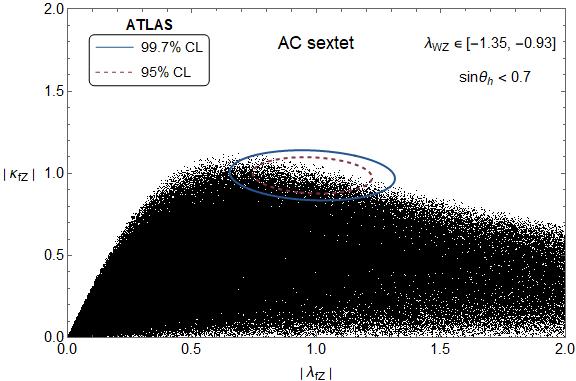}     
    \caption{Updated constraints considering the correct $\kappa_{f}$ and including the strict $2 \sigma$ bound from $b\rightarrow s\gamma$~\cite{GGM}.
    }
    \label{fig:neglamACC}
\end{figure}

For AC pentets and AC sextets, the situation is similar and the analysis becomes model-dependent because it depends on the size of the vevs and the correlation between $\vec{R}$ and the vevs. We can see the difference in the Figure~\ref{fig:neglamACC}, where we impose the strict bound from $b\rightarrow s\gamma$~\cite{GGM} assuming that all additional scalars are heavy. This bound can be translated to $\sin\theta_{h} <0.7$, where $\sin\theta_{h}$ is the fraction from the extended representations to the EW vev:
\begin{align}
 \sin \theta_H= \Biggl\{\begin{array}{ll}
\sqrt{8} v_\chi / v & \text { AC triplet } \\
\sqrt{20} v_4 / v & \text { AC quartet } \\
\sqrt{40} v_5 / v & \text { AC pentet } \\
\sqrt{70} v_6 / v & \text { AC sextet }
\end{array} \, .
\end{align} 
We can see that both the AC pentet and AC sextet now have points inside the $2\sigma$ region. If the model-independent constraints get stronger, then the allowed region from the scan will shrink and it may become possible to extend the exclusion to the AC pentet and AC sextet with our methodology.

The inclusion of small custodial violation allowed by the uncertainty in the measurement of the $\rho$ parameter makes only a small difference. Additionally, the case with multiple AC triplets is still excluded, since the discussion of the individual contributions from each multiplet still holds.

\bibliographystyle{apsrev-title}
\bibliography{BIBNEGLWZ}

\end{document}